\documentclass[a4paper,11pt]{article}

\usepackage{amsmath,amsfonts,amssymb,theorem,verbatim}
\usepackage{amsfonts,amssymb,graphics,theorem}
\usepackage[greek, english]{babel}
\usepackage[pdftex]{graphicx}

\usepackage{float,multirow}
\floatstyle{ruled}
\newfloat{algo}{thp}{lop}
\floatname{algo}{Algorithm}

\newtheorem{Thm}{Theorem}
\newtheorem{Def}{Definition}
 \newtheorem{Lemma}{Lemma}
\newtheorem{Prop}{Proposition} 

\def\BEN{\begin{enumerate}}  \def\BI{\begin{itemize}}
\def\EEN{\end{enumerate}}   \def\EI{\end{itemize}} 
   
  \def\no{\noindent}

\def\nn{\nonumber}
\def\beq{\begin{eqnarray}} \def\eeq{\end{eqnarray}}

\def\al*#1{\begin{align*}#1\end{align*}}

\def\ga*#1{\begin{gather*}#1\end{gather*}}

\def\alat*#1#2{\begin{alignat*}{#1}#2\end{alignat*}}
\def\bea{\begin{eqnarray*}}
\def\eea{\end{eqnarray*}}
\def\ml*#1{\begin{multline*}#1\end{multline*}}

 \def\mbf{\mathbf} \def\mrm{\mathrm}
 \def\unl{\underline} \def\ovl{\overline}
\newcommand{\Bf}[1]{{\mbox{\scriptsize\boldmath$#1$}}}
\newcommand{\bff}[1]{{\mbox{\boldmath$#1$}}}

\def\le{\left} \def\ri{\right}

\def\te#1{\mathrm{e}^{#1}}  \def\td{\text{\rm d}}

\def\i{\infty}

\def\T{\tilde}  
\def\WH{\widehat} \def\WT{\widetilde}

\def\a{\alpha}

\def\m{\mu} 
  \def\nn{\nonumber}   \def\s{\sigma}
     
  \def\q{\qquad}

\newcommand{\exit}{{\mbox{\, \vspace{3mm}}} \hfill\mbox{$\square$}}

\begin{document}

\title{Pricing and hedging barrier options\\
in a hyper-exponential additive model}
\author{ Marc Jeannin$^{\dagger, \star}$ and Martijn Pistorius$^{\star}$\\
\phantom{..............}\\
$\phantom{|}^{\dagger}${\small Models and Methodology Group,
Risk Management Department}\\
{\small Nomura International plc}\\
{\small Nomura House 1 St Martin's-le-Grand, London EC1A 4NP, UK}\\
{\small E-mail: marc.jeannin@nomura.com}\\
\phantom{..............}\\
$\phantom{|}^{\star}${\small Department of Mathematics, Imperial
College London}\\
{\small South Kensington Campus, London SW7 2AZ, UK}\\
{\small E-mail: martijn.pistorius@imperial.ac.uk }\\
\phantom{..............}\\}
\date{}

\maketitle
\begin{abstract}
In this paper we develop an algorithm to calculate the prices and
Greeks of barrier options in a hyper-exponential
additive model with piecewise constant parameters.
We obtain an explicit semi-analytical expression for
the first-passage probability.
The solution rests on a randomization
and an explicit matrix Wiener-Hopf factorization.
Employing this result we derive
explicit expressions for the Laplace-Fourier transforms of the
prices and Greeks of barrier options.
As a numerical
illustration, the prices and Greeks of down-and-in digital and
down-and-in call options are calculated for a set of parameters
obtained by a simultaneous calibration to Stoxx50E
call options across strikes and four different maturities.
By comparing the results with Monte-Carlo simulations, we show that the method
is fast, accurate, and stable.
\bigskip

\textit{Keywords:} Hyper-exponential additive processes, matrix
Wiener-Hopf factorization, first passage times, barrier options, multi-dimensional
Laplace transform, Fourier transform, sensitivities.
 \vfill

\textit{Acknowledgements:}  We would like to thank P.~Howard and
S.~Obraztsov for their support, and also D.~Madan for useful
conversations. This research was supported by EPSRC grant
EP/D039053, and was partly carried out while the authors were
based at King's College London.

\end{abstract}

\thispagestyle{empty}
\newpage{}
\setcounter{page}{1}
\section{Introduction}

Barrier options are contracts whose pay-offs are activated or
de-activated when the underlying process crosses a pre-specified
level. These contracts are among the most popular path-dependent options. To
value barrier options, a model needs to be sufficiently flexible
to calibrate call option prices at different strikes and maturities.
However, it is desirable to maintain a degree of
analytical tractability to facilitate the calculations, especially
for the Greeks or the sensitivities. These sensitivities describe the change in the model price with respect to a change in the underlying parameter, and are important for an appreciation of the
robustness of the model's results. It is well known that the accurate evaluation of the
Greeks is a challenging numerical problem, since standard PDE or
Monte-Carlo methods are generally slow and unstable.

It is well established that the geometric Brownian motion
model lacks the flexibility to capture features in financial asset
return data such as the skewness and the excess kurtosis. It cannot
calibrate simultaneously to a set of call option prices. To address these limitations,
one of the approaches consists of introducing jumps
in the price process by replacing the Brownian motion by a
L\'{e}vy process. L\'{e}vy models, such as the VG, CGMY, NIG,
KoBoL, generalised hyperbolic, and Kou's double
exponential model, have been successfully applied to the valuation
of European-type options. We refer to Cont and Tankov
\cite{ContTankov}, Boyarchenko and Levendorskii \cite{BLbook}, and
Schoutens \cite{Schoutens} for background and references on the
application of L\'{e}vy models in option pricing.

As observed by many authors, such as Eberlein and Kluge \cite{EberleinKluge}, or
Carr and Wu \cite{carr03}, L\'{e}vy
models are generally not capable of calibrating option prices
simultaneously across strikes and maturities. Empirical studies of S\&P500 index data by
Carr and Wu \cite{carr03}, and Pan \cite{pan02}, show that the implied jump intensities and the implied jump size distributions vary greatly over time. The prices of short-dated options exhibit a significantly larger risk-premium than that of long-dated options. This is reflected in the thicker tails of the implied marginal risk-neutral distributions, especially at short maturities. For example, in the equity markets, short-dated
out-of-the money put options are relatively expensive since the risk of a
large negative jump in the share is priced.  Because of the stationarity and independence of the increments of a L\'{e}vy process, the moments exhibit a rigid
term structure that is different from what is observed in market data.
This lack of flexibility can be overcome by considering models driven by additive processes,
which have independent and time-inhomogeneous increments.

Additive models have been used for equity option pricing by
Carr et al.~\cite{carr06}, Galloway and Nolder \cite{galloway04}, and by Eberlein and Kluge \cite{EberleinKluge} for interest rate option pricing.
Motivated by modelling considerations, Carr \cite{carr06} proposed a
self-similar additive model for the log-price, and reported good calibration results
across time. Galloway and Nolder \cite{galloway04} carried out a calibration study for various related models. Eberlein and Kluge \cite{EberleinKluge} constructed an HJM model driven by an additive process with continuous characteristics, and they obtained a good fit for swaptions by using piecewise constant parameters.

In this paper we follow a similar approach: we model the share price by an
additive process with hyper-exponential jumps.
Hyper-exponential distributions are finite mean-mixtures of
exponential distributions which can approximate monotone distribution arbitrarily closely.
As first observed by Asmussen et al.~\cite{AMP}, most of the popular L\'{e}vy models
used in mathematical finance possess completely monotone L\'{e}vy densities
and can therefore be approximated well by hyper-exponential L\'{e}vy models. A hyper-exponential additive model is sufficiently flexible to allow for an accurate calibration
to European option prices across strikes and multiple maturities.
In addition, if the parameters are piecewise
constant, the model admits semi-analytical expressions for
prices and Greeks of barrier options.


There currently is a body of literature devoted to various
aspects of pricing barrier options. In the setting
of L\'{e}vy models, a transform-based approach to price barrier
options has been developed in a number of papers, including
Geman and Yor \cite{GemanYor},
Kou and Wang \cite{KouWang}, Davydov and Linetsky \cite{DavydovLinetsky},
Boyarchenko and Levendorskii \cite{BoyLev}.
In particular, Kou and Wang \cite{KouWang}, Kou et al.~\cite{KouPW},
Sepp \cite{Sepp}, Lipton \cite{Lipton}, and Jeannin and Pistorius \cite{JeanninP} considered the cases of L\'{e}vy processes with double-exponential and
hyper-exponential jumps.


In this paper, the transform algorithm that we develop is based on a so-called matrix
Wiener-Hopf factorization. Such matrix factorizations were
first studied by London et al.~\cite{london80} and Rogers
\cite{rogers94} for (noisy) fluid models. Jiang and Pistorius
\cite{JiangP} developed matrix-Wiener factorization results for
regime-switching models with jumps. We show that by suitably
randomizing the parameters the distributions of
the infimum and supremum of the randomized hyper-exponential additive process
can be explicitly expressed in terms of a matrix Wiener-Hopf
factorization. We use these results to derive semi-analytical
expressions for the first-passage time probabilities,
for the prices, and for the Greeks of barrier options,
up to a multi-dimensional transform.
The actual prices are
subsequently obtained by inverting this transform.

As a numerical illustration, we calibrate the hyper-exponential
additive model to Eurostoxx prices quoted on 27 February 2007 at four different maturities.
We calculate in this setting down-and-in digital and down-and-in
call option prices and Greeks (delta and gamma).
To invert the transform, we use a contour deformation
algorithm and a fractional Fast Fourier Transform algorithm, developed
by Talbot \cite{talbot79}, Bailey and Swarztrauber \cite{Bailey94}, and
Chourdakis \cite{choudhury94}, \cite{chourdakis04}.
We also compare it to Monte-Carlo Euler scheme simulations. We find that the
algorithm is accurate and stable, and much faster than
Monte-Carlo simulations (especially for the Greeks).
This method is suitable for applications in which the number of
periods is not too large (up to four). When a larger number of periods is required,
the direct inversion method used here is no longer feasible.
The subject still needs to be further investigated and is left for future research.

The remainder of the paper is organized as follows. In Section
\ref{sec:add} we define the hyper-exponential additive model and
present its application to European call option pricing. In
Sections \ref{sec:WH} and \ref{sec:price} we derive
semi-analytical expressions for the first-passage probabilities of
a hyper-exponential additive process in terms of a matrix
Wiener-Hopf factorisation, and for the prices and Greeks of
barrier options. In Section \ref{sec:num} we
present numerical results.

\section{The model}\label{sec:add}
\subsection{Additive processes}
We consider an asset price process $S$ modelled as the exponential
$$ S_t = S_0 \te{X_t}$$ of an additive process $X$. Informally, an
additive process can be described as a L\'{e}vy process with
time-dependent characteristics or, equivalently, as a process with
independent but non-stationary increments. We briefly review below
some key properties of additive processes. For further background on additive processes and their applications in finance, we refer to Sato \cite{Sato}, and to Cont and Tankov \cite{ContTankov}.
An additive process can be defined more formally as follows.

\begin{Def}\label{def:locallevy}
For a given $T>0$, $X = \{X_t, t\in[0,T]\}$ is an additive process if
\begin{itemize}
\item[(i)] $X_0=0$, \item[(ii)] For any finite partition $0\leq
t_0<t_1 \cdots <t_k\leq T$, the random variables $X_{t_k}-X_{t_{k-1}}, \cdots, X_{t_0}$ are
independent, \item[(iii)] The sample paths $t\mapsto X_t$ have
c\`adl\`ag modifications almost surely.
\end{itemize}
\end{Def}
If $X$ is an additive process, then, for every $t\in[0,T]$, $X_t$
has an infinitely divisible distribution with L\'{e}vy triplet
$(M_t, \Sigma^2_t, \Lambda_t)$; that is, the characteristic
function of $X_t$ is given by $\Phi_t(u) = \exp
[\Psi_{t}(u)\,]$. According to the L\'{e}vy-Khintchine
formula, $\Psi_{t}$ is the characteristic
exponent given by
$$
\Psi_{t}(u) = \mathbf i u M_t - \frac{\Sigma^2_t}{2} u^2 +
\int_{-\i}^\i \le\{\te{\mathbf iux} - 1 - \mathbf
iux1_{\{|x|<1\}}\ri\}\Lambda_t(\td x),
$$
with $M_t$, $\Sigma_t\in\mathbb R$, and where $\Lambda_t$ the
L\'{e}vy measure satisfies the integrability constraint $$\int
(1\wedge x^2)\Lambda_t(dx)<\i.$$
The law of the additive process $\{X_t, t\in[0,T]\}$ is determined by the collection of L\'{e}vy triplets
$\{(M_{t},\Sigma^2_{t}, \Lambda_{t})\ \text{for}\ t\in[0,T]\}$. If
the L\'{e}vy triplets are time-independent, $X$ is a L\'{e}vy process.
If the additive process has absolutely continuous characteristics,
the L\'{e}vy triplets take the explicit form
\begin{eqnarray*}
M(t) &=& \int_0^t \mu(s)\td s, \qquad
\Sigma(t) = \int_0^t\sigma^2(s)\td s,\\
\Lambda_t(B) &=& \int_0^t\int_B g(s,x)\td x\td s
\qquad \text{for Borel sets $B$},
\end{eqnarray*}
where $\mu,\sigma^2:[0,T]\to\mathbb R$
and $g:[0,T]\times\mathbb R\to\mathbb R$
are integrable functions, with $g$ and $\sigma^2$
non-negative. We call the functions $(\mu,\sigma^2, g)$
the local triplet of $X$.\\
\\
We assume that we have been given
deterministic integrable functions $r(t)$ and
$d(t)$ representing the short rate and the
dividend yield, and that
the characteristic exponent of $X_t$ satisfies
\begin{equation}\label{eq:psi1}
\Psi_t(-\mathbf i) = \int_0^t[r(s)-d(s)]\td s,\qquad
\end{equation}
or equivalently
\begin{equation}\label{eq:psi2}
\mu(t) + \frac{\sigma^2(t)}{2} +
\int_{-\infty}^\infty [\te{x} - 1 - 1_{\{|x|<1\}}x]g(t,x)\td x
= r(t) - d(t).
\end{equation}
It follows that the discounted process
$$
 S_{0}\exp(\int_0^t [r(s)-d(s)]\td s)
$$
is a martingale if and only if
\eqref{eq:psi1} (or, equivalently, \eqref{eq:psi2}) holds.

%
%
%
%
\subsection{Hyper-exponential additive processes}
In what follows we restrict the discussion to a hyper-exponential additive
process $X$ which is specified by its local triplet
$(\mu,\sigma^2,g)$ where $g$ is given by
$$
g(t,x) =
\sum_{k=1}^{n^+}\pi^+_k(t)\alpha^+_k(t)\te{-\alpha^+_k(t)x}1_{\{x>0\}} +
\sum_{j=1}^{n^-}\pi^-_j(t)\alpha_j^-(t)\te{-\alpha^-_j(t)|x|}1_{\{x<0\}},
$$
where $\pi_k^\pm(t)$ and $\alpha^\pm_k(t)$ are non-negative. The
continuous part of $X$ consists of a diffusion with time-dependent
drift $\mu(t)$ and volatility $\s(t)$. The jump part of $X$
is of finite activity and forms an inhomogeneous
compound Poisson process where positive and negative jumps occur
at the rates
$$\lambda^+(t):=\sum_{k=1}^{n^+}\pi^+_k(t) \q\text{ and }\q
\lambda^-(t):= \sum_{j=1}^{n^-}\pi^-_j(t),$$ and jump sizes
are distributed according to a hyper-exponential distribution.\\
Small random price movements are intuitively modelled by
the diffusion part, whereas sudden changes of the price
are captured by the jump-part of $X$. If we take $n^\pm = 1$,
the jump-sizes are exponentially distributed, and this model reduces to an extension of
the Kou model with time-dependent parameters.

\subsection{Piecewise constant parameters}
To reduce the dimension of the available parameter set,
we take the functions $\mu(t),\sigma(t)$ and $g(t,\cdot)$ to be piecewise constant.
Given that we have a finite set of European call options
with different maturities $T_1, \ldots, T_N$, we take the
local parameters to be constant between the different maturities $T_i$.
Then for all $t \in (T_{i-1},T_i]$, (with $T_0=0$) we set
\begin{equation}\label{eq:para}
\mu(t) = \mu^{(i)},\quad \sigma^2(t) = \sigma^{2(i)},\quad g(t,x)
= g^{(i)}(x), \quad\text{$i=1, \ldots, N$}.
\end{equation}
For $t\in(T_{i-1}, T_i]$ the characteristic exponent of
$X_t-X_{T_{i-1}}$ is  given by $$\Psi_{T_{i-1},t}(u) =:
\Psi^{(i)}(u),$$ where
\begin{equation}\label{eq:Psi}
\Psi^{(i)}(u) = \mu^{(i)} u\mathbf i - \frac{\sigma^{2(i)}}{2}u^2
+ \sum_{k=1}^{n^+} \pi_k^{+(i)}\le(\frac{u\mathbf i}{\alpha_k^{+(i)} -
u\mathbf i}\ri) - \sum_{j=1}^{n^-} \pi_j^{-(i)}\le(\frac{u\mathbf
i}{\alpha_j^{-(i)} + u\mathbf i}\ri).
\end{equation}

\section{First passage probabilities}\label{sec:WH}
The value of a digital barrier option  can be expressed in terms of
the distribution
$$
F^{(+)}(x;T) = P(\ovl X_T \leq x)
$$
of the running supremum
$$
\ovl X_T = \sup_{s\leq T} X_s
$$
of $X$, or equivalently, the distribution of the
first-passage time
$$
T^+(x)=\inf\{t\ge0: X_t>x\}
$$
which is related to $F^{(+)}$ by
$$
P(T^+(x)\leq T) = 1-F^{(+)}(x;T).
$$
Whereas for a L\'{e}vy process the distributions of the infimum
and supremum are linked to the characteristic exponent
by the so-called Wiener-Hopf factorization, such a result does not
exist for general additive processes, because of the
time-dependence of the parameters. However, in the case of
piecewise constant parameters, the triplet changes only at
deterministic times, so that as a consequence
the distribution function $F^{(+)}(x) = F^{(+)}(x; T^{(1)},
\ldots, T^{(N)})$ only depends on the inter-jump times $T^{(i)} =
T_i - T_{i-1}$ (with $T_0=0$). In this case, as we show below,
the $N$-dimensional
Laplace transform $G^{(+)}(x,\mbf q)$ of $F^{(+)}$, given by
$$ G^{(+)}(x,\mbf q) = \int \te{- (q_1 u_1 + \cdots\, + q_Nu_N)}
F^{(+)}(x; u_1, \ldots, u_N)\td u_1 \cdots \td u_N,
$$
where $\mbf q = (q_1, \ldots, q_N)$,
is expressed explicitly in terms of a matrix Wiener-Hopf
factorization. To state this result we need to introduce some
further notation.

For any vector $\mbf v=(v_1, \ldots, v_n)$, we denote by $\Delta_{\Bf v}$ the diagonal matrix
$\Delta_{\Bf v} = (v_i, i=1, \ldots, n)_{\mrm{diag}}$.
Let $Q$ be the $N(1+n^++n^-)\times N(1+n^++n^-)$ matrix given in
block notation by
\begin{equation}\label{eq:Q}
Q = \begin{pmatrix}   H^+ & D^-\\ C^- & T^-
\end{pmatrix},
\end{equation}
where
\begin{equation}\label{eq:H}
H^+ = \begin{pmatrix}   G - \Delta_{\Bf\lambda} & b^+\\
t^+ & T^+
\end{pmatrix}.
\end{equation}
Here $\bff\lambda=(\lambda_i^+ + \lambda_i^-, i=1, \ldots, N)$,
and $G$ and $b^+$ are the $N\times N$ and $N\times Nn^+$ matrices in
block notation given by
\begin{equation}\label{eq:b+}
G = \begin{pmatrix} -q_1 &q_1 &&&\\
 & -q_2&q_2&&\\
&&\ddots& \\&&&-q_{N-1} & q_{N-1}\\
&&&& -q_N
\end{pmatrix},
\
b^+ = \begin{pmatrix} \mbf\pi^{+{(1)}} &&&\\
 & \mbf\pi^{+{(2)}} &&\\
&&\ddots&\\  &&& \mbf\pi^{+{(N)}}
\end{pmatrix}.
\end{equation}
Here $\mbf\pi^{+(i)}$ is the row-vector $\bff\pi^{+(i)}=
(\pi^{+(i)}_l, l=1, \ldots, n)$, and where $t^+$, $T^+$ are given by
\begin{align}\label{eq:t+}
T^+ &=  \begin{pmatrix} -\Delta_{\Bf\a^+} &&&\\
 & -\Delta_{\Bf\a^+} &&\\
&&\ddots&\\  &&& -\Delta_{\Bf\a^+}
\end{pmatrix},\quad
t^+ = \begin{pmatrix} \mbf\a^+  &&&\\
 & \mbf\a^+ &&\\
&&\ddots&\\  &&& \mbf\a^+
\end{pmatrix},
\end{align}
where $\mbf\a^+$ is the column-vector $\mbf\a^+ = (\alpha_i^+,
i=1, \ldots, n^+)'$, and
\begin{equation}\label{eq:C-D-}
C^-=
\begin{pmatrix} t^-& O^- \end{pmatrix},
\quad\quad\quad D^- = \begin{pmatrix} b^- \\ O^+ \end{pmatrix},
\end{equation}
where $O^\pm$ are $n^\pm N\times n^\pm N$ zero matrices, and
$b^-$, $T^-$, and $t^-$ are given by \eqref{eq:b+} and
\eqref{eq:t+} with $\bff\pi^{+(i)}$ and $\bff\alpha^+$ replaced by
$\bff\pi^{-(i)}$ and $\bff\alpha^-$. The matrix $Q$ is a generator matrix, that is,
a square matrix with non-negative off-diagonal elements and
non-positive row sums, and defines a Markov
chain. This Markov chain is associated to a randomization and
embedding of the additive process $X$ (which will be illustrated
with a concrete example below). We recall that a
sub-probability matrix is a matrix with non-negative elements and
row sums not larger than one. By applying the matrix Wiener-Hopf
factorization results of Jiang and Pistorius \cite{JiangP} to the current
setting we arrive at the following conclusion.

\begin{Thm}\label{thm:wh} It holds that
\begin{equation}\label{eq:GX}
G^{(+)}(x,\bff q) = \frac{1}{q_1 \ldots q_N} \times \left[ 1 - e_1'
\te{Q_+ x}\mathbf 1\right],
\end{equation}
where
$$
\mbf e_1'=(1, 0, \ldots, 0)\q\text{ and }\q\mathbf 1=(1,\ldots, 1)'.
$$
$Q_+$ is an $N(1+n^+)\times N(1+n^+)$ generator
matrix that together with $\eta^+$ an $Nn^-\times N(1+n^+)$
sub-probability matrix, solves the system of matrix
equations
\begin{equation}\label{eq:WH+}
 \begin{cases}
 \frac{1}{2}S^2 Q_+^2 - V^+ Q_+ + H^+ + D^-\eta^+ = O,\\
\\
-\eta^+ Q_+ + C^- + T^-\eta^+ = O.
 \end{cases}
\end{equation}
Here the $O$'s are zero matrices of appropriate sizes, and in block
notation we have,
\begin{equation}\label{eq:TSV}
S^2 = \begin{pmatrix}   \Delta_{\Bf\s^2} & \\ & O^+
\end{pmatrix},
\qquad V^+ = \begin{pmatrix} + \Delta_{\Bf\m} & \\ & I^+
\end{pmatrix},
\end{equation}
with
$$\bff\s^2=(\s_i^2, i=1,\ldots, N),\qquad \bff\mu=(\mu_i, i=1,
\ldots,N).$$ $O^+$ and $I^+$ represent $n^+N\times n^+N$
zero and identity matrices, respectively.
\end{Thm}
By applying Theorem \ref{thm:wh} to $-X$, we find the
corresponding pair of matrices $(Q_-,\eta_-)$. The quadruple
$(Q_+,\eta_+,Q_-, \eta_-)$ is called a matrix Wiener-Hopf
factorization of $Q$.\\
\\
{\bf Example.} To illustrate this approach, we consider a
hyper-exponential additive process $X$ on $[0,T_2]$ whose
parameters are constant during the periods $[0,T_1]$ and
$[T_1, T_2]$. In the first period $X$ evolves as a jump-diffusion
with positive and negative exponential jumps with means and jump
rates $1/\alpha^+, \lambda^+$ and $1/\alpha^-, \lambda^-$. In
the second period $X$ is a Brownian motion with drift. The idea is to
randomize the times between maturities by replacing $T^{(1)}=T_1$
and $T^{(2)}=T_2-T_1$ with independent exponential random variables
having means $q_1^{-1}$ and $q_2^{-1}$. This results in a
regime-switching jump-diffusion with the regime only jumping from
state 1 to state 2, according to the generator matrix
$$
G = \begin{pmatrix}
  -q_1 & q_1\\ 0 & -q_2
\end{pmatrix}.
$$
We associate to the regime-switching process a continuous
Markov additive process, which can be informally obtained by
replacing positive and negative jumps with stretched slopes of
$+1$ and $-1$ (see Asmussen \cite{asm00} for background on this
embedding). As described in \cite{JiangP}, in this case the
generator of the modulating Markov process is given by
$$
Q = \begin{pmatrix}   H^+ & D^-\\ C^- & T^-
\end{pmatrix}
=
 \le(\begin{array}{ccc|c}   -q_1-\lambda^+-\lambda^- & q_1
& \lambda^+ &
\lambda^-\\
0 & -q_2 & 0 & 0\\
\alpha^+ & 0 & -\alpha^+ & 0\\
\hline \alpha^- & 0 & 0 & -\alpha^-
\end{array}\ri),
$$
with the matrices $S^2$ and $V^+$ in Theorem \ref{thm:wh} given by
$$
S^2 = \begin{pmatrix}   \sigma_1^2 \\
& \sigma_2^2\\
&& 0
\end{pmatrix},
\qquad V^+ =  \begin{pmatrix}   \mu_1 \\
& \mu_2\\
&& 1
\end{pmatrix}.
$$

\begin{figure}[htbp]
\centering
        \includegraphics[height=10.4cm,width=7.55cm]{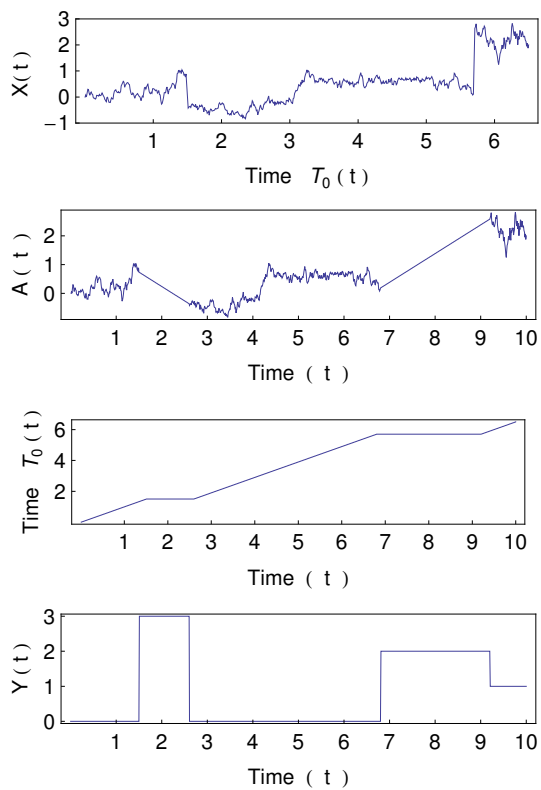}
\caption{Paths of the various processes related to log-price process $X$ are illustrated.
Here $X$ is a hyper-exponential additive process on the period $[0,T_2]$
whose parameters are constant during the periods $[0,T_1]$ and
$[T_1, T_2]$. During the first period $[0,T_1]$, the process
$X$ evolves as a jump-diffusion with volatility $\sigma_1$, drift $\mu_1=0$ and
exponentially-distributed jumps. The positive and negative jumps are
exponentially-distributed with means and jump rates $1/\alpha^+, \lambda^+$
and $1/\alpha^-, \lambda^-$, respectively. During the second period $[T_1,T_2]$, $X$
evolves as a linear Brownian motion with volatility $\sigma_2$ and drift $\mu_2=0$.
Associated to $X$ is a continuous Markov additive process $A$, which
can be obtained from $X$ by replacing the positive and negative jumps
with linear stretches of path of slopes $+1$ and $-1$, and replacing the fixed times
$T_1$, $T_2-T_1$ by independent exponential random times $\mbf e_1$ and $\mbf e_2$ with parameters $q_1, q_2$. The process $Y$ that records the current state or regime of $A$ is a Markov process with generating matrix $Q$.  When $Y$ takes values 1 and 2, $A$ evolves as a linear Brownian motion with zero drift and volatility $\sigma_1$ and $\sigma_2$ respectively; and when $Y$ is 3 and 4, $A$ is a positive or negative unit drift. These linear stretches of paths of $A$ originate from the jumps of $X$. A jump of $Y$ from one state to another is induced either by a jump of $X$ or by a switch of the set of parameters that determine the dynamics of $X$. By time-changing $A$ by the time $T_0(t)$ up to time $t$ that $Y$ was equal to $1$ and $2$, we recover a regime-switching jump-diffusion; that is, the process $\{A(T_0(t)), t\ge 0\}$ is in law equal to a regime-switching jump-diffusion. Finally, replacement of the times at which a regime switch occurs by $T_1$ and $T_2$ results in  a process that has the same distribution as $X$.}\label{Fig:emb}
\end{figure}

\subsection{Solution of the matrix equation}
To solve the system \eqref{eq:WH+}, which is a Ricatti-type matrix
equation, we follow a spectral approach and determine the
spectral decomposition of $Q^+$. Denoting by $h(\rho)$ a (column)
eigenvector of $Q^+$ corresponding to the eigenvalue $\rho$, one finds that it is a
matter of algebra to verify that the system \eqref{eq:WH+} can be
equivalently rewritten as
$$
\frac{1}{2} \T S^2\begin{pmatrix}I\\\eta^+\end{pmatrix}Q_+^2 - \T
V\begin{pmatrix}I\\\eta^+\end{pmatrix}Q_+ +
Q\begin{pmatrix}I\\\eta^+\end{pmatrix} = O.
$$
Here $O$ is a $N(1+n^+ + n^-)$ square zero matrix, $I$ is an
$N(1+n^+)$ identity matrix, and
$$
\T S^2 = \begin{pmatrix}   \Delta_{\Bf\s^2} & \\ & O^+\\ && O^-
\end{pmatrix},
\qquad \T V = \begin{pmatrix}   \Delta_{\Bf\m} & \\ & I^+ \\ &&
-I^-
\end{pmatrix}.
$$
Defining the matrix $K(s)$ by
\begin{equation}\label{eq:K}
K(s) = \frac{s^2}{2}\T S^2 + s\T V + Q,
\end{equation}
we find that $h(\rho)$ solves the linear system
$$K[-\rho]\begin{pmatrix}I\\\eta^+\end{pmatrix}h(\rho)=\mbf 0,$$
which implies that $\rho$ is a root of the equation $\det K(s) =
0$.
The following result characterizes the eigenvalues of $Q^+$
(see Appendix A):
\begin{Lemma}\label{lem:eigen}\quad\quad
\begin{itemize}
\item[(i)] It holds that
\begin{equation}\label{eq:detK}
|\det (K(s))| = \prod_{i=1}^N \le\{|\Psi^{(i)}(-\mathbf i s) -
q_i| \prod_{k=1}^{n^+} |s-\alpha^+_k| \prod_{l=1}^{n^-}
|s+\alpha^-_l|\ri\},
\end{equation}
where $\Psi^{(i)}$ is given in \eqref{eq:Psi}.
\item[(ii)] The equation \begin{equation}\label{eq:detK2} \det
K(s)=0\end{equation} has $N(1+n^+)$ positive roots and $N(1+n^-)$
negative roots.
\end{itemize}
\end{Lemma}
Since $-Q^+$ is the negative of a generator matrix, it is
non-negative definite, so its eigenvalues are non-negative
and are given by the positive roots of \eqref{eq:detK2}. In
particular, if the positive roots $$\bff\rho_+=\le(\rho^+_i: i=1,
\ldots, N(n^++1)\ri)$$ of equation \eqref{eq:detK2} are distinct, it
follows from Lemma \ref{lem:eigen} and Theorem \ref{thm:wh} that
$$ G^{(+)}(x,\mbf q) = (q_1\cdots q_N)^{-1} \times
[1-\mbf e_i' U_+ \te{-\Delta_{\Bf\rho_+} x} U_+^{-1}\mathbf 1],$$ where
$U_+ = (h(\rho^+_i), i=1, \ldots, N(n^++1))$.

\subsection{The final position and the first exit time}
The valuation of barrier options involves the joint distribution
of the final position at maturity $T$ and the first exit time. We
will extend the results in the previous section by considering the following:
\begin{eqnarray*}
\ovl F^{(+)}(x,s) &:=& E[\te{sX_T}\mathbf 1_{\{\ovl X_T > x\}}]\\
&=& E[\te{sX_T}\mathbf 1_{\{ T^+(x) < T\}}].
\end{eqnarray*}
$\ovl F^{(+)}(x,s)$ depends on time only
through the inter-maturity times $(T^{(1)},\ldots, T^{(N)})$.
The Laplace transform $\ovl G^{(+)}(x,s;q)$ of $\ovl
F^{(+)}(x,s)$ in $(T^{(1)},\ldots, T^{(N)})$, can be
expressed in terms of $Q^+$ and $K(s)$ as follows:
\begin{Prop}\label{prop:XS} It holds that
\begin{equation}\label{eq:GXS}
  \ovl G^{(+)}(x,s,\mbf q) = \frac{\te{sx}}{q_1\ldots q_N}
  \times \mbf e_1' \te{Q^+ x}K(s)^{-1}
  K(0)\mathbf 1
\end{equation}
for all $s\in\mathbb C$ with $\mathrm{Re}(s)\in (-\min_{j=1,\ldots, n^-}
\a_j^-,\min_{k=1,\ldots,n^+} \a^+_k)$.
\end{Prop}
A proof is given in Appendix A.

\subsection{First passage to a lower level}\label{ssec:low}
The form of the analogous distributions concerning the infimum
$$
F^{(-)}(x) = P(- \unl X_T\leq x), \qquad \ovl F^{(-)}(x,s) =
E[\te{sX_T}\mathbf 1_{\{- \unl X_T > x\}}],\quad x>0,
$$ can be found by
applying the results in the previous section to the process $-X$.
More specifically, it is straightforward to check that the
$N$-dimensional Laplace transforms $G^{(-)}(x,q)$ and $\ovl
G^{(-)}(x,s,q)$ are given by \eqref{eq:GX} and \eqref{eq:GXS} replacing
$Q^+$ by $Q^-$. $(Q^-,\eta^-)$ satisfies the system
of matrix equations \eqref{eq:WH+} with $V^+, H^+, D^-, C^-, T^-$
replaced by $V^-, H^-, D^+, C^+, T^+$, where the latter set is
defined by interchanging $+$ and $-$ in equations \eqref{eq:TSV},
\eqref{eq:C-D-}, \eqref{eq:t+} and \eqref{eq:H}.
It is straightforward to verify that \textit{(i)} an eigenvector
$h(\rho)$ of $Q^-$ corresponding to eigenvalue $\rho$ satisfies
$$
K[\rho]\begin{pmatrix}\eta^-\\ I\end{pmatrix}h(\rho)=0,
$$
where $I$ is an $N(1+n^-)$ identity matrix, and \textit{(ii)} that, in view of
Lemma \ref{lem:eigen}, the eigenvalues of $Q^-$ are given by the
negative roots
$$
\bff\rho^- = \le(\rho^-_j, j=1, \ldots, N(n^-+1)\ri)
$$
of $\det K(\rho) = 0$.

\section{Prices and Greeks of digital and barrier
options}\label{sec:price} Using the first-passage results from
the previous section we derive semi-analytical expressions
for the prices and sensitivities of down-and-in digital and
knock-in call options. A down-and-in digital option at level
$H<S_0$ is a contract that pays out one unit at maturity $T$ if the
price $S$ has down-crossed the level $H$ before $T$. Similarly, a down-and-in
call option at level $H<S_0$ and with strike $K$ is a call option
whose pay-off is activated once $S$ down-crosses $H$. Taking the
risk-free rate $r$ and the dividend rate $d$ to be constant,
the arbitrage free prices of a down-and-in digital $(DID)$ and a call option $(DIC)$ are given respectively by
\begin{equation*}
  DID(T,H,S_0) = \te{-(r-d)T} E\le[\mathbf 1_{\{\inf_{s\leq T} S_s < H\}}\ri]
  = \te{-(r-d)T} P(\unl X_T < h),
\end{equation*}
where $h=\log(H/S_0)$ is the log-barrier, and
\begin{equation*}
  DIC(T,H,K,S_0) 
  = \te{-(r-d)T} S_0 E\le[(\te{X_T} - \te{k})^+\mathbf 1_{\{\unl X_T <
  h\}}\ri],
\end{equation*}
where $k=\log(K/S_0)$ denotes the log-strike.
Let $\WH{DID}(\mbf q)$ denote the joint Laplace transform of $DID$ in
the inter-maturity times $(T^{(1)}, \ldots, T^{(N)})$ (with
$T^{(N)}=T$), and denote by $\WH{DIC}^*(\mbf q,s)$ the Laplace-Fourier
transform in $(T^{(1)}, \ldots, T^{(N)})$ and in the log-strike $k$.
Then we have the following result:

\begin{Prop}\label{prop:price} For $h=\log(H/S_0)<0$ it holds that
\begin{eqnarray}\label{eq:DID}
\WH{DID}(\mbf q) &=& \frac{1}{c(q)} \times \mbf e_1' \te{Q^- h}\mathbf 1,\\
\WH{DIC}^*(\mbf q,s) &=& \frac{S_0\te{bk}}{c(\mbf q) b(b-1)} \times
\mbf e_1'
\te{Q^- h}K(s)^{-1}K(0)\mathbf 1, \label{eq:DIC}
\end{eqnarray}
where $k=\log(K/S_0)$, $\mbf q=(q_1,\ldots, q_N)$, and $b=\alpha +
\mathbf i s + 1, c(\mbf q) = (q_1 + r) \cdots (q_N + r).$
\end{Prop}

Before we give the proof we observe that from the explicit
expressions \eqref{eq:DID} and \eqref{eq:DIC} semi-analytical
formulas can be obtained for the delta and gamma of the
down-and-in digital and call options (i.e.~the first and second
derivatives of the option value with respect to the spot $S_0$).
Indeed, the derivatives of the expressions \eqref{eq:DID} and
\eqref{eq:DIC} with respect to $S_0$ are equal to the
Laplace-Fourier transforms of the derivatives of the option, as
integration and differentiation are interchangeable in this case.
In the case of a down-and-in digital option we find
that the Laplace transforms $\WH{\Delta}_{DID}$ and
$\WH{\Gamma}_{DID}$ of the delta $\Delta_{DID}$ and gamma
$\Gamma_{DID}$ are given by
$$
\WH{\Delta}_{DID}(\mbf q) = -\frac{1}{c(q)S_0}\times e_1' Q^-\te{Q^-
h}\mathbf 1, \ \quad \WH{\Gamma}_{DID}(\mbf q) =
\frac{1}{c(q)S_0^2}\times e_1' [(Q^-)^2 + Q^-]\te{Q^- h}\mathbf 1.
$$

\no{\it Proof of Proposition \ref{prop:price}:} The expression
\eqref{eq:DID} is a direct consequence of Theorem \ref{thm:wh}
(see also Section \ref{ssec:low}). To verify \eqref{eq:DIC} we
start by taking the Fourier transform in $k$ and find as in
\eqref{eq:FFTCall} that the Fourier transform $DIC^*$ is given by
\begin{equation}\label{eq:DIC*}
DIC^*(\alpha+\mathbf i s) = \frac{\ovl F^{(-)}(-h,b)}{b(b-1)},
\end{equation}
where $b = \alpha+\mathbf i s + 1$ and
$$
\ovl F^{(-)}(x,b) = E[\te{bX_T}\textbf{1}_{\{-\unl X_T \geq x\}}].
$$
From Proposition \ref{prop:XS} we deduce that the form the joint
Laplace transform $\ovl G(x,b,\mbf q)$ of $\ovl F^{(-)}(x,b)$ in
$(T^{(1)},\ldots, T^{(N)})$ is given by
\begin{equation}\label{eq:ovlG}
\ovl G(x,b,\mbf q) = \frac{\te{bx}}{c(\mbf q)} \times \mbf e_1' \te{Q^-
h}K(s)^{-1}K(0)\mathbf 1.
\end{equation}
Combining \eqref{eq:DIC*} and \eqref{eq:ovlG} completes the
proof.\exit

\section{Numerical results}\label{sec:num}
\subsection{Calibration}
To determine a parameter set to test the method,
we calibrate the hyper-exponential additive model
to Eurostoxx call options at four
different maturities, observed in the market on 20 February
2007. The spot price is EUR $4150$, the risk-free rate is
assumed to be fixed at $r=0.03$, and the dividend rate is taken to
be zero. As we find that inclusion of positive jumps does not
substantially improve the calibration results, we only consider
negative jumps, and we specify the jump size parameters to
be $(\alpha^-_1, \alpha^-_2)=(3,10)$.
The jump arrival rates \mbox{\boldmath$\pi$$^{-}$} and the volatility are
piecewise constant in time, and are estimated by minimizing the
root-mean-square error between model and observed market call
prices. Using the well known Fourier transform method (briefly recalled in Appendix \ref{app:euro}) the calibration is carried out maturity by maturity under
constraints through a bootstrapping method with well-defined local triplets:

\begin{enumerate}
\item Calibrate call prices at $T_1$ to obtain the parameters
$(\sigma(T_1),\pi_1^{\pm}(T_1),\pi_2^{\pm}(T_1))$. \item For $j=2,
\ldots, N$ calibrate call prices at $T_j$ to obtain
$(\sigma(T_j),\pi_1^{\pm}(T_j),\pi_2^{\pm}(T_j))$.
\end{enumerate}

In Figure 2 the calibration results are presented with
plots of the market and model implied volatility surfaces
corresponding to the four maturities $6$m, $1$Y, $3$Y and $5$Y.
The root-mean-square error (RMSE) and
the average relative percentage error (ARPE) are equal
to $5.30$ and $1.1\%$. We compare it to a price process
that follows a L\'{e}vy process with
hyper-exponential jumps  (i.e.~with constant
parameters over time), and find that the calibration of
the four maturities in that case
give a RMSE of $9.82$ and an ARPE of $2.9\%$. In Table 1 the
resulting parameter sets are displayed under the hyper-exponential additive
and L\'{e}vy models.
In the case of the hyper-exponential model we
observe a high jump intensity of small jumps
for short maturities that decrease substantially over time.
This is consistent with
the finding of Carr and Wu \cite{carr03},
and Pan \cite{pan02}.

\begin{figure}[p]\label{fig:cal}
\centering
        \includegraphics[scale =0.71]{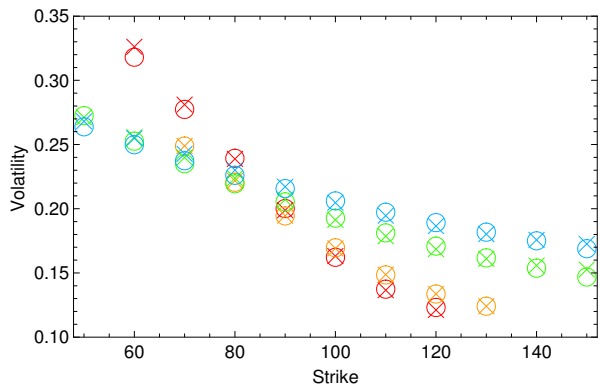}
\caption{Calibration of Eurostoxx call option prices on February
20th 2007 for an additive hyper-exponential model with piecewise
constant parameters. Crosses represent market implied volatilities
and circles model implied volatilities for four maturities: 6 months
(red), 1 year (orange), 3 years (green), 5 years (blue). }
\end{figure}
\begin{table}[p]
\centering {\small
\begin{tabular}{|cc|ccc|}
\hline
\textbf{Additive hyperexponential model}&&&&\\
\hline
$T_{Start}$& $T_{End}$&$\sigma$ & $\pi^-_1$ & $\pi^-_2$ \\
\hline
$0$&$0.5$&$0.0995$&$0.0371$&$11.1819$\\
$0.5$&$1$&$0.0759$&$0.2091$&$9.9540$\\
$1$&$3$&$0.0786$&$0.4738$&$7.0322$\\
$3$&$5$&$0.0858$&$0.8084$&$0.2361$\\
\hline \hline
\textbf{L\'{e}vy hyperexponential model}&&&&\\
\hline
$T_{Start}$& $T_{End}$&$\sigma$ & $\pi^-_1$ & $\pi^-_2$ \\
\hline
$0$&$5$&$0.1171$&$0.5693$&$0.0165$\\
\hline
\end{tabular}
\caption{Calibration of Eurostoxx call option prices quoted on
February 20 2007 for an additive hyper-exponential model with
piecewise constant parameters and for a L\'{e}vy hyper-exponential
model with jump parameters $\alpha_1^-=3$ and
$\alpha^-_2=10$. The interest and dividend rates are $r=0.03$ and
$d=0$. The maturities are given in years.} \label{table:LocalLevy}  }
\end{table}

\subsection{Results for the barrier option prices and Greeks}
Using the parameter set found in the calibration of the Eurostoxx call options,
we value barrier and digital options on the Eurostoxx index,
modelling its price process as the exponential of a
hyper-exponential additive process.
We use the semi-analytical results in
Proposition \ref{prop:price}. To invert the multi-dimensional
Laplace transforms we choose Talbot's method \cite{talbot79}
(see also \cite{choudhury94}) for down-and-in digital options. We
combine it with the fractional FFT algorithm of Bailey and Swarztrauber \cite{Bailey94}, and Chourdakis \cite{chourdakis04} for down-and-in call options. See Appendix \ref{app:trans}
for a detailed description of the implementation of these transform algorithms.
We compare it to the same quantities calculated by Monte-Carlo simulations, using a standard
Euler scheme.

\subsubsection{Down-and-in digital options}
We price down-and-in digital options with a maturity of five years for
different spot levels. We evaluate the required
4-dimensional Laplace transform over two time increments of
six months, and two of two years, that is,
$T^{(1)}=0.5$, $T^{(2)}=0.5$, $T^{(3)}=2$, $T^{(4)}=2$.
We use Talbot's algorithm with $M=6$ (see Appendix \ref{app:trans}
for an explanation of this parameter). Using Mathematica to run
the algorithm, the computation time was five minutes on a 3189 Mhz
computer to calculate prices and Greeks for fourteen different spot
levels. The calculation of first passage probabilities using Monte-Carlo simulations requires a large number of time steps and paths.
We use one million paths with $\delta t= 5 \times 10^{-5}$ and it
takes several hours to obtain stable Greeks in C++. Error bounds
cannot be obtained analytically, but we observe in Table
\ref{table:VGDD} that the results of the transform method agree
with Monte-Carlo simulation results. Figures 3 and 4 report prices
and Greeks for down-and-in digital options. The options
are expressed as a percentage of the spot price.
The values of the sensitivities are expressed as fractions of the spot price $S_0$.

This transform algorithm is particulary efficient at a book level,
since once the generating matrices $Q^-$ of the infimum have been
calculated for different values of the vector $\textbf{q}$ the
calculation of prices and Greeks of any digital barrier product
is just a matter of summation.

\subsubsection{Down-and-in call options}
We value down-and-in call options with a maturity of one year for
different strike levels. In this case, a
two-dimensional Laplace inversion is required over time increments
$T^{(1)}=0.5$ and $T^{(2)}=0.5$. For the inversions of the Laplace
transform and the Fourier transform, we set $M=7$ and
$N=1024$ (refer to Appendix \ref{app:trans}
for an explanation of these parameters). For Monte-Carlo simulations, we use one million
paths with time step $\delta t= 2.5 \times 10^{-5}$. The option
prices and Greeks obtained by the two methods are reported in
table \ref{table:VGDIC} and figures \ref{fig:DID5YG} and \ref{fig:DIC1Y}.
We observe that the results of the transform
method agree with the Monte-Carlo simulation results. Using
Mathematica again to run the algorithm, the computation time is ten minutes
to calculate the option prices, delta and gamma for eleven different
levels of the strike. Since the option prices and Greeks of a
down-and-in call option are obtained via a Fourier-Laplace
transform, it takes more time than
in the case of a digital option (approximately twice as long),
which is still much faster than a Monte-Carlo Euler scheme. We
note that the transform algorithm is particulary efficient for the
pricing of options with different strikes, as we obtain by FrFFT inversion
the prices and Greeks of any down-and-in call options on a log-strike grid.

\newpage
\begin{figure}[htbp]
\centering
        \includegraphics[scale =0.85]{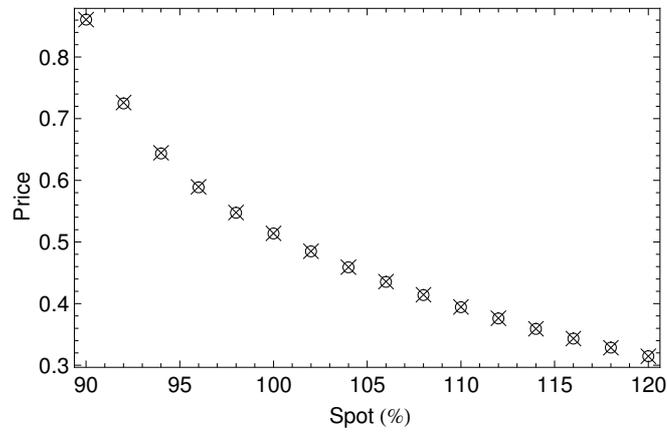}
\caption{\small Prices of down-and-in digital options with a
barrier $H$ set at $90\%$ of EUR $4150$ and maturity $T=5$ years.
Semi-analytical results are indicated with the symbol $\times$ and Monte-Carlo
results with the symbol $\circ$. } \label{fig:DID5Y}
\end{figure}

\bigskip
\bigskip
\bigskip
\bigskip

\begin{figure}[htbp]
\centering
    \hspace{-0.8cm}
        \includegraphics[scale =0.65]{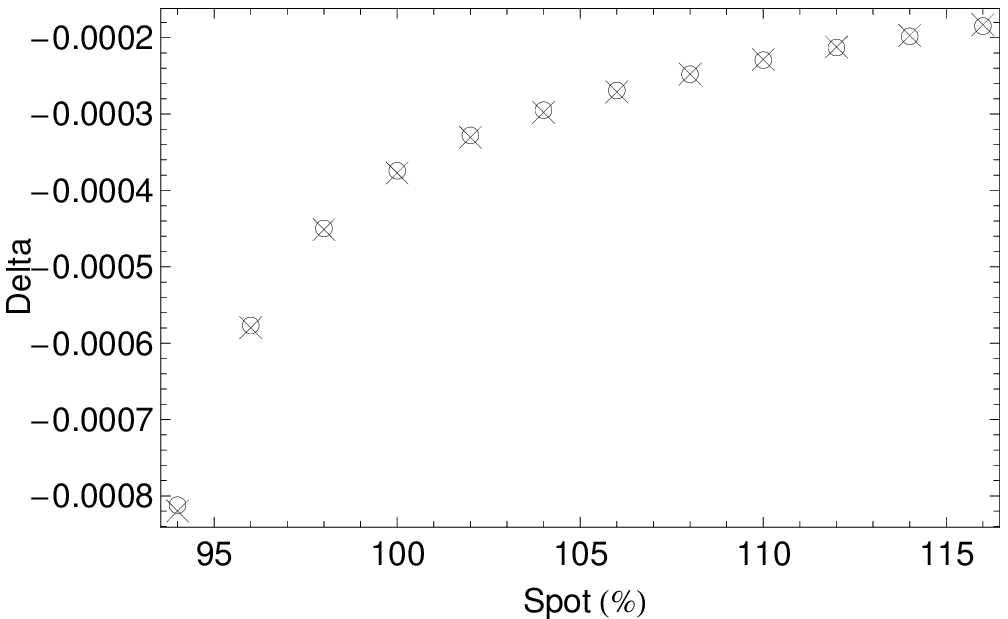}
        \hspace{+0.7cm}
                \includegraphics[scale =0.65]{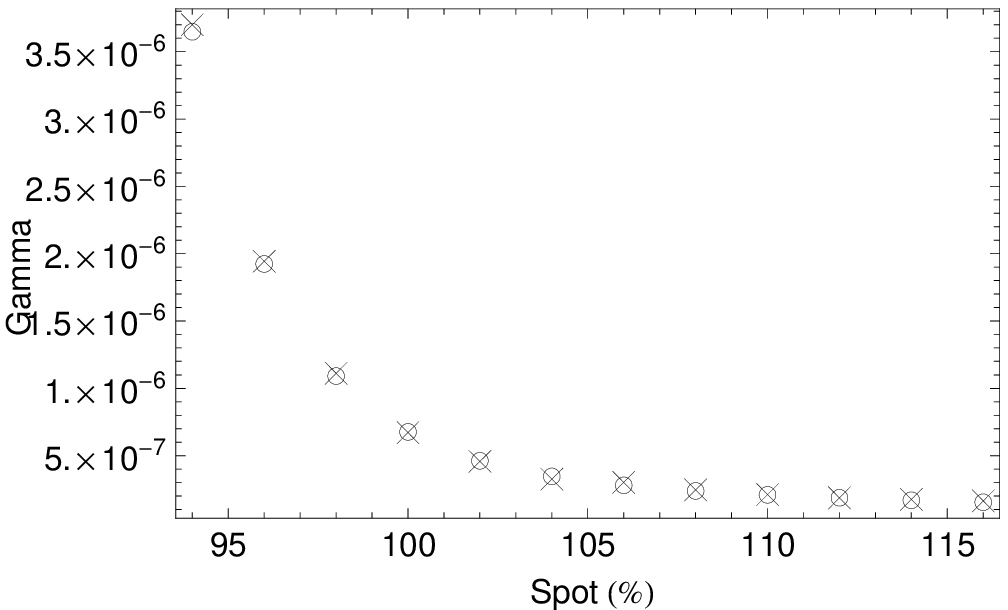}\\
\caption{\small Greeks of down-and-in digital options with a
barrier $H$ set at $90\%$ of EUR $4150$ and maturity $T=5$ years.
Semi-analytical results are indicated with the symbol $\times$ and
Monte-Carlo results with the symbol $\circ$. } \label{fig:DID5YG}
\end{figure}

\begin{figure}[htbp]
\centering
        \includegraphics[scale =0.65]{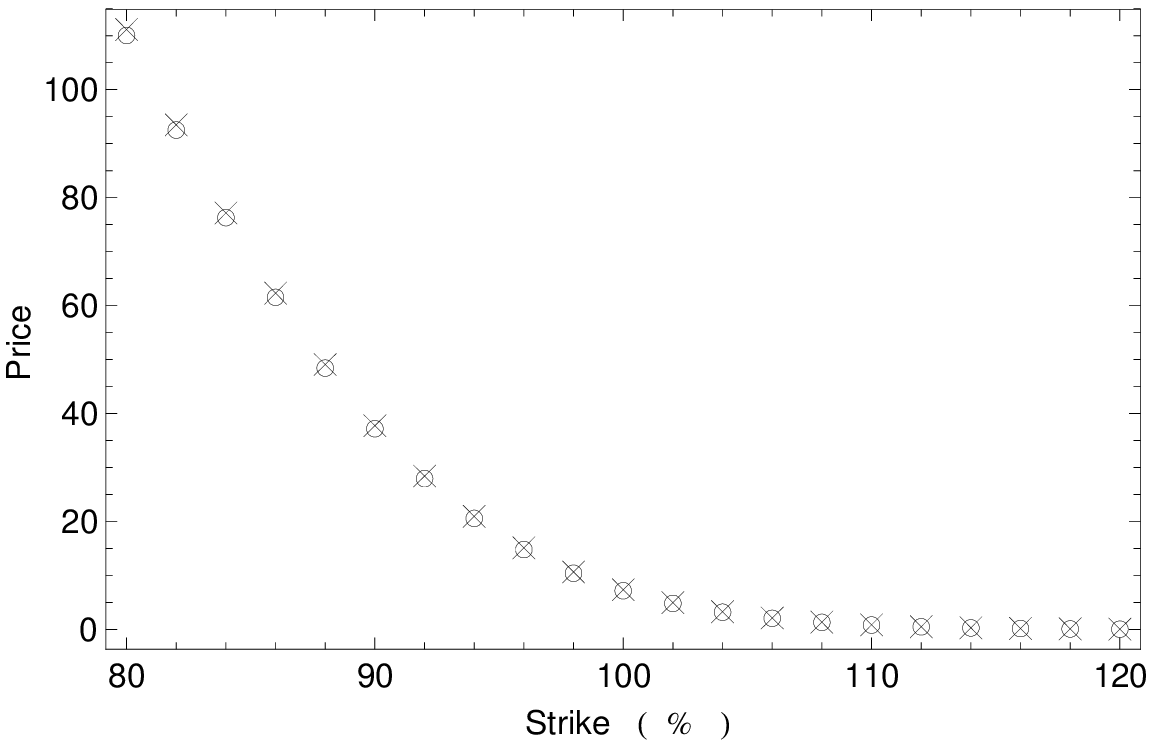}
\caption{\small Prices of down-and-in call options with a barrier
$H$ set at $90\%$, a spot at EUR $4150$ and maturity $T=1$ year.
Semi-analytical results are indicated with the symbol $\times$ and
Monte-Carlo results with the symbol $\circ$. } \label{fig:DIC1Y}
\end{figure}

\begin{figure}[htbp]
\centering
    \hspace{-0.8cm}
        \includegraphics[scale =0.65]{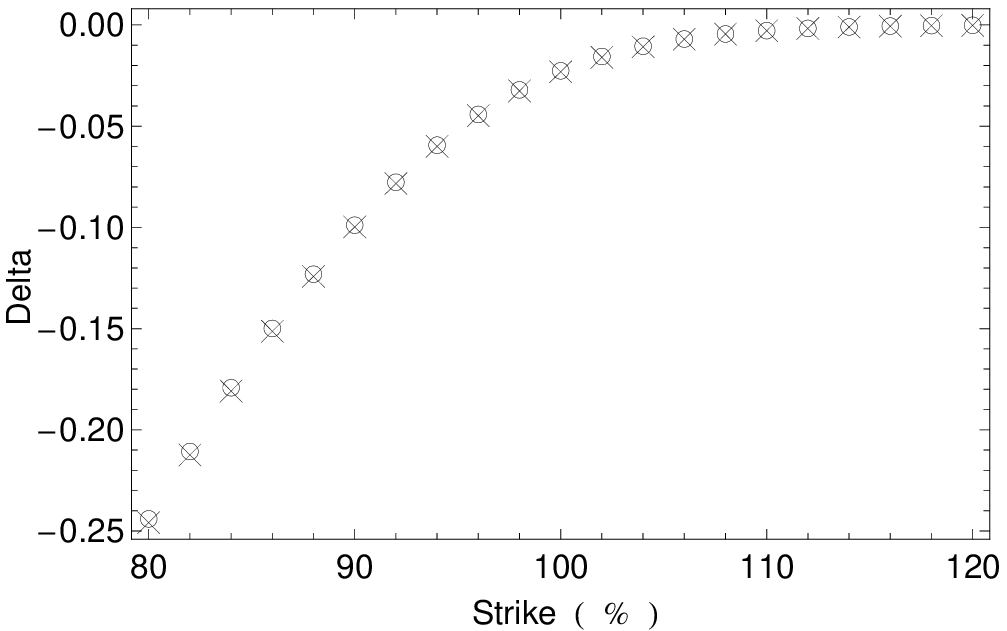}
        \hspace{+0.7cm}
                \includegraphics[scale =0.65]{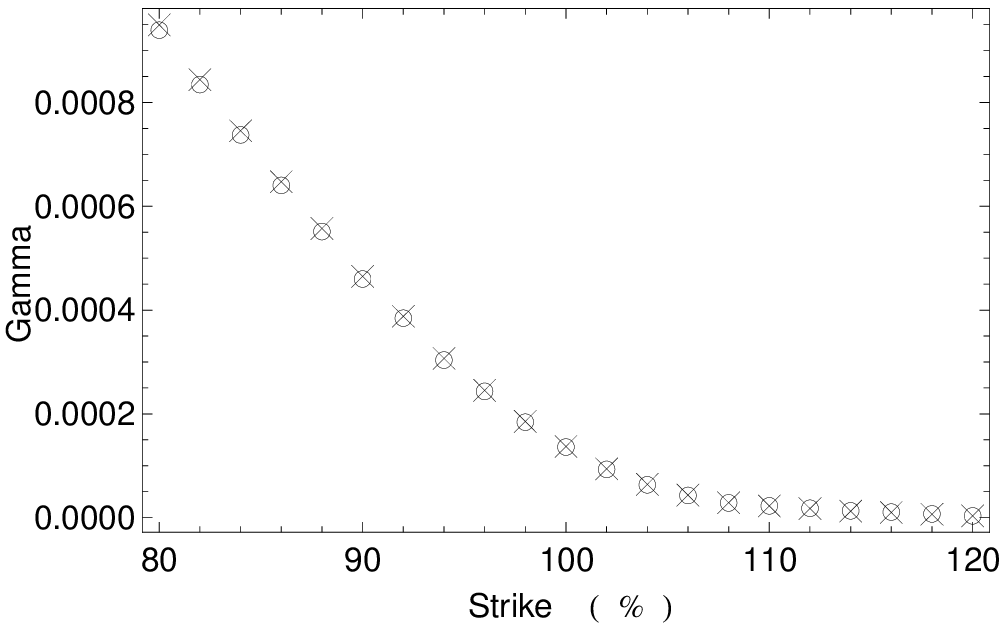}\\
\caption{\small Greeks of down-and-in call options with a barrier
$H$ set at $90\%$, a spot at EUR $4150$ and maturity $T=1$ year.
Semi-analytical results are indicated with the symbol $\times$ and Monte-Carlo
 results with the symbol $\circ$. } \label{fig:DIC1YG}
\end{figure}

\newpage
\begin{table}[htbp]
\centering {\tiny
\begin{tabular}{|r||rr|rr|rr|}
\hline
&\multicolumn{6}{|c|}{Down-and-in digital options} \\
\hline
 &\multicolumn{2}{|c|}{Price}&\multicolumn{2}{|c|}{Delta $(\times 10^{-3})$}&
\multicolumn{2}{|c|}{Gamma $(\times 10^{-7})$}\\
\hline
 $\%$   & TA    & (MC$-$, MC$+$) & TA & MC  & TA &
MC \\
\hline
92& 0.7261&(    0.7000, 0.7499)&    -1.226&  -1.232&  7.34&    7.54\\
94& 0.6446&(    0.6137, 0.6735)&    -0.812&  -0.819&  3.64&    3.70\\
96& 0.5893&(    0.5566, 0.6207)&    -0.577&  -0.579&  1.92&    1.94\\
98& 0.5478&(    0.5143, 0.5805)&    -0.450&  -0.451&  1.09&    1.11\\
100&    0.5140&(    0.4800, 0.5475)&    -0.374&  -0.377&  0.67&    0.67\\
102&    0.4850&(    0.4506, 0.5189)&    -0.328&  -0.330&  0.46&    0.45\\
104&    0.4592&(    0.4245, 0.4932)&    -0.294&  -0.297&  0.34&    0.33\\
106&    0.4358&(    0.4008, 0.4697)&    -0.269&  -0.271&  0.28&    0.30\\
108&    0.4144&(    0.3795, 0.4483)&    -0.247&  -0.247&  0.24&    0.24\\
110&    0.3946&(    0.3599, 0.4285)&    -0.229&  -0.228&  0.21&    0.21\\
112&    0.3762&(    0.3518, 0.4010)&    -0.212&  -0.212&  0.19&    0.19\\
114&    0.3592&(    0.3250, 0.3917)&    -0.198&  -0.197&  0.17&    0.17\\
116&    0.3433&(    0.3095, 0.3771)&    -0.184&  -0.182&  0.15&    0.16\\
118&    0.3285&(    0.3012, 0.3632)&    -0.172&  -0.170&  0.14&    0.14\\
\hline
\end{tabular}
\caption{\small Down-and-in digital options with barrier $H$ set at $90\%$ of EUR 4150.
The first column contains the spot price as
a percentage figure of $4150$.  The columns with TA contain the
results obtained using the transform algorithm, whereas MC refers
to Monte-Carlo results. In the case of the price the Monte-Carlo
results are reported in the form of a $95\%$ confidence interval,
and in the other cases as a point estimate.} \label{table:VGDD} }
\end{table}
%
%
%
\begin{table}[htp]
\centering {\tiny
\begin{tabular}{|r||rr|rr|rr|}
\hline
&\multicolumn{6}{|c|}{Down-and-in call options} \\
\hline
 &\multicolumn{2}{|c|}{Price}&\multicolumn{2}{|c|}{Delta $(\times 10^{-1})$}&
\multicolumn{2}{|c|}{Gamma $(\times 10^{-4})$}\\
\hline
 $\%$   & TA    & (MC$-$ , MC$+$) & TA & MC  & TA  &
MC\\
\hline
80& 111.13&(    108.85, 111.20)&    -2.458&  -2.440&  9.49&    9.39\\
82& 93.48&(    91.46, 93.55)&    -2.124&  -2.107&  8.43&    8.34\\
84& 77.14&(    75.35, 77.19)&    -1.808&  -1.792&  7.45&    7.37\\
86& 62.28&(    60.71, 62.33)&    -1.514&  -1.499&  6.47&    6.40\\
88& 49.08&(    47.71, 49.13)&    -1.242&  -1.231&  5.57&    5.51\\
90& 37.63&(    36.7, 37.80)&    -0.991&  -0.988&   4.65&    4.59\\
92& 28.11&(    27.43, 28.46)&    -0.783&  -0.776&  3.88&    3.84\\
94& 20.94&(    20.13, 21.01)&    -0.599&  -0.593&  3.07&    3.03\\
96& 15.11&(    14.45, 15.16)&    -0.448&  -0.443&  2.45&    2.43\\
98& 10.66&(    10.12, 10.72)&    -0.326&  -0.320&  1.85&    1.84\\
100& 7.36&(   6.92, 7.41)&    -0.231&  -0.226&  1.37&    1.36\\
102& 4.97&(   4.63, 5.02)&    -0.159&  -0.156&  0.94&    0.93\\
104& 3.28&(   3.03, 3.34)&    -0.107&  -0.104&  0.64&    0.63\\
106& 2.12&(   1.94, 2.18)&    -0.070&  -0.069&  0.43&    0.42\\
108& 1.35&(   1.22, 1.41)&    -0.045&  -0.044&  0.28&    0.28\\
110& 0.84&(    0.74, 0.89)&    -0.028&  -0.027&  0.23&    0.22\\
112& 0.51&(    0.47, 0.54)&    -0.017&  -0.017&  0.18&    0.18\\
114& 0.30&(    0.25, 0.34)&    -0.010&  -0.009&  0.13&    0.13\\
116& 0.18&(    0.16, 0.19)&    -0.006&  -0.006&  0.10&    0.10\\
118& 0.10&(    0.08, 0.11)&    -0.003&  -0.003&  0.07&    0.07\\
120& 0.06&(    0.03, 0.06)&    -0.002&  -0.002&  0.03&    0.03\\
\hline
\end{tabular}
\caption{\small Down-and-in call options with a barrier $H$ set
at $90\%$, a spot at EUR $4150$ and maturity $T=1$ year for a range of
strikes. The first column contains the strike level as a percentage
figure of the spot EUR $4150$.  The columns with TA contain the
results obtained using the transform algorithm, whereas MC refers
to Monte-Carlo results. In the case of the price the Monte-Carlo
results are reported in the form of a $95\%$ confidence interval,
and in the other cases as a point estimate.} \label{table:VGDIC} }
\end{table}

\newpage{}
\appendix

\centerline{\bf APPENDIX}
\section{Proofs}

{\it Proof of Lemma \ref{lem:eigen}:} \begin{itemize}
\item[(i)] It is straightforward to
verify that $K^\#(s)$ can be obtained from $K(s)$ by interchanging
some columns and rows, where $K^\#(s)$ is given by
\[ K^\#(s) =\left( \begin{array}{ccccc}
K_1(s) & D_{1} &&&\\
 & K_2(s) &D_{2}& &\\
&&.&.&\\
&&& K_{N-1}(s) &D_{N-1} \\
 && & &K_N(s)\\
\end{array} \right),\]
where $K_w(s)$ and $D_w$ are square matrices of dimension
$n^++n^-+1$. There are defined respectively by
\[ K_w(s) = \left( \begin{array}{ccccccc}
\frac{\sigma_w^2s^2}{2} + \mu_ws-q_w-\lambda_w^+-\lambda_w^-
&\pi_1^{-}(w)&.
&\pi_m^{-}(w)&\pi_1^{+}(w)&.&\pi_n^{+}(w)\\
\alpha_1^-&-s-\alpha_1^-&0&0&0&0&0\\
.&0&.&0&0&0&0\\
\alpha_m^-& 0 &0&-s-\alpha_m^-&0&0&0\\
\alpha_1^+&0&0&0&s-\alpha_1^+&0&0\\
.&0&0&0&0&.&0\\
\alpha_n^+&0&0&0&0&0&s-\alpha_n^+
\end{array} \right),\]
where $\lambda_w^\pm=\sum_i \pi^\pm_i(w)$, and
$$
(D_w)_{ij} =
\begin{cases}
q_w & \text{if $i=j=1$}\\ 0 & \text{otherwise}
\end{cases}.
$$
Therefore $|\det(K(s))|$ is equal to $|\det(K^\#(s)|$. To proceed
we recall an identity from matrix algebra. Let $M$ be a matrix of
the form
$$
M = \begin{pmatrix}   M_{11} & M_{12} \\ M_{21} & M_{22}
\end{pmatrix}
$$
in block notation, where $M_{22}$ is invertible. Then
$$
\det(M) = \det(M_{22}) \det (M_{11} + M_{12} M_{22}^{-1} M_{21}).
$$
Using this identity, it is a matter of algebra to
verify by induction that
$$
\det(K^\#(s)) = \det(K_1(s)) \cdots det(K_N(s)).
$$
As a consequence we find, by applying this matrix identity,
that
\begin{multline*}
\hspace{-0.5cm}\det(K_w(s)) = \le(\frac{\sigma^2_w}{2} s^2 + \mu_w s +
\sum_{i=1}^{n^+} \pi^+_i(w) \frac{\alpha_i^+}{s-\alpha_i^+} +
\sum_{j=1}^{n^-} \pi^-_j(w) \frac{\alpha_j^-}{-s-\alpha_j^-}
-\lambda_w^+ - \lambda_w^- - q_w\ri)\times \\
\times\prod_{i=1}^{n^+} (s-\alpha^+_i) \prod_{j=1}^{n^-}
(-s-\alpha^-_j),
\end{multline*}
and the assertion follows in view of \eqref{eq:Psi}.

\item[(ii)]  Using the intermediate value theorem and the specific form of
$\Psi$, it is straightforward to check that the equation
$\Psi(-u\mathbf i)=q$, $q>0$ has $n+1$ positive roots $\rho^+_i$
and $m+1$ negative roots $\rho^-_j$, satisfying
$$
\rho^-_{m+1} < -\alpha_m^- < \rho_m^- < \ldots < -\alpha_1^- <
\rho^-_1<0 < \rho_i^+ < \alpha_1^+ < \ldots < \rho_n^+ <
\alpha_n^+ < \rho^+_{n+1}.
$$
Since $\det(K_w(s))$ is a polynomial of degree $n+m+2$, it follows
that all the roots of $\det(K_w(s))=0$ are given by $(\rho_i^+,
i=1, \ldots, n+1)$ and $(\rho_j^-, j=1,\ldots, m+1)$. In view of
the form of $\det(K(s))$ derived in (i) the assertion follows.
\end{itemize}
\exit
\\
\\
{\it Proof of Proposition \ref{prop:XS}:} Consider the following
randomization of $X$ obtained by randomizing the inter-maturity
times $T^{(i)}$ by replacing them by independent exponential
random times with means $q_i^{-1}$, and call this process $\WT X$.
The process $\WT X$ is a regime-switching jump-diffusion, where
the only regime switches that can occur are from $i$ to $i+1$ at
rate $q_i$ ($i=1,\ldots, N-1$), and from the final state $N$ to an
absorbing 'graveyard state' $\partial$. As shown in \cite{JiangP},
the process $\WT X$ is equal to a time-changed continuous process
$A$, say. Denoting by $\zeta$ the epoch at which $A$ is sent to
$\partial$, by $Y$ the modulating Markov chain, and by
$\tau=\inf\{t\ge0: A_\tau=x\}$, we have
\begin{eqnarray*}
  q_1\cdots q_N \ovl G(x,s,q) &=& E[\te{s A_{\zeta-}} \mathbf 1_{\{\ovl A_{\zeta-} >
  x\}}]\\
&=& E[\te{s A_{\zeta-}} \mathbf 1_{\{\tau < \zeta\}}]\\
&=& E[\te{s A_{\tau}} \mathbf 1_{\{\tau < \zeta\}}f(Y_\tau)]\\
&=& \te{s x} E[\mathbf 1_{\{\tau < \zeta\}}f(Y_\tau)],
\end{eqnarray*}
with
$$
f(y) = E[\te{s A_{\zeta-}}|A_0=0, Y_0=y],
$$
where the last two lines follow by the Markov property of $(A,Y)$
and the fact that $A$ is continuous.  To guarantee that all the
expressions are well defined in this calculation $s$ has to be
such that $E[\te{s X_1}]<\i$, which corresponds to the restriction
that $$Re(s)\in(-\min_j \a_j^-,\min_i \a^+_i).$$

In \cite{JiangP} it was shown that the vector $\mbf f = (f(y),
y\in N)$, where $N$ denotes the state space of $Y$, is given by
$$
\mbf f =  K(s)^{-1} Q \mathbf 1,
$$
where the matrix $K(s)$ is given in (\ref{eq:K}). Combining these
results with Theorem \ref{thm:wh} we find that
$$
q_1\cdots q_N \ovl G(x,s,q) = e_1' \te{Q^+ x} K(s)^{-1} Q \mathbf
1,
$$
and the proof is complete. \exit

\section{European call options}\label{app:euro}
Under the hyper-exponential
additive model with piecewise constant parameters \eqref{eq:para},
the characteristic function at time $T$ is explicitly given by
$$\Phi^{(i)}(u) = \exp\le(\sum_{j=1}^i\Psi^{(j)}(u)\ri),$$
with $\Psi^{(j)}$ as given in \eqref{eq:Psi}. The price of a European
call with maturity $T_i$ can thus be efficiently calculated using a
well-established Fourier transform method, which we briefly recall. The Fourier
transform $C^*_{T_i}$ over $k$ of $C_{T_i}(k)$, the price of a call option
with log-strike $k=\log(K/S_0)$ and maturity $T_i$, can be explicitly
expressed in terms of the characteristic function $\Phi^{(i)}(u)$
as follows:
\begin{eqnarray}
\nn C^*_{T_i}(v-\mathbf i \alpha)&=&S_0 \te{-rT_i}
\int_{-\infty}^{\infty}\te{\mathbf i vk}E[\te{\alpha k}(\te{X_{T_i}}-\te k)^+] \td k\\
&=&S_0\te{-rT_i}\frac{\Phi^{(i)}(v-(\alpha+1)\mathbf
i)}{(\alpha+\mathbf iv)(\alpha+1+\mathbf iv)}.\label{eq:FFTCall}
\end{eqnarray}
Since the call pay-off function itself is not square-integrable in
the log-strike, the axis of integration is here shifted over
$\mathbf i\alpha$ which corresponds to exponentially dampening the
pay-off function at a rate $\alpha$, which is usually taken to be
$\alpha=0.75$ (see Carr and Madan \cite{carr98}). The call option
prices are then determined by inverting the Fourier transform:
\begin{equation}\label{eq:CallTi}
C_{T_i}(k)=\frac{S_0 \te{-\alpha k}\te{-rT_i}}{\pi}\int_0^{\infty}
\te{-\mathbf ikv}\frac{\Phi^{(i)}(v-(\alpha+1)\mathbf
i)}{(\alpha+\mathbf iv)(\alpha+1+\mathbf iv)}\td v.
\end{equation}

\section{Transform inversion algorithms}\label{app:trans}

\subsection{Multi-dimensional Laplace inversion}
To evaluate down-and-in digital option prices (DID), we
invert the multi-dimensional Laplace transform \eqref{eq:DID} to obtain
\begin{equation}\label{eq:lDID}
DID(S_0,h,\textbf{T})=\frac{1}{(2\pi \mathbf
i)^N}\int_{C_N}\cdots\int_{C_1}\te{q_1 T_1+\ldots+q_NT_N}
\WH{DID}(S_0,h,\textbf{q})\td \textbf{q}.
\end{equation}
where $\mbf T=(T_1, \ldots, T_N)$ and
$C_n$ are vertical lines in the complex plane defined by
$q_n=r_n+\mathbf iy_n$ for $n=1, \ldots N$ with
$-\infty<y_n<\infty$ and fixed values of $r_n$, chosen such
that all the singularities of the transform
$\WH{DID}(S_0,h,\texttt{q})$ are coordinate-wise on the left of
the lines $C_n$. Many algorithms approximate the integrals in
\eqref{eq:lDID} by a finite linear combination of the transform
at some specific nodes with certain weights. Three approaches
have been studied by Abate et al.~\cite{abate06}, based on Fourier series expansion,
combinations of Gaver functionals, and deformation of the integral contour. Here we
concentrate on the last method developed by Talbot \cite{talbot79}, since
reports in the literature (e.g. \cite{abate06}) suggest that
this approach offers high performance for a short time of
execution, which our numerical results confirm. We write
\begin{equation}\label{eq:DIDLT}
DID(S_0,h,\textbf{T})=\frac{1}{(2\pi \mathbf
i)^N}\int_{-\pi}^{\pi}\cdots\int_{-\pi}^{\pi} \beta_1(\theta)
\cdots \beta_N(\theta) \WH{DID}(S_0,h,\textbf{q}(\theta)) \td
\mbox{\boldmath $\theta$},
\end{equation}
with $n=1, \ldots N$, $\beta_n(\theta)=w_n \te{\mathbf i r_n w_n T_n}$, $ q_n(\theta)=\mathbf i r_n w_n$, and
$$
w_n=-1+\mathbf i \theta+\mathbf i (\theta \cot \theta-1) \cot \theta.
$$
Since $DID$ is a real valued function, $DID$ is also equal to the real part of
the integral on the right-hand side of \eqref{eq:DIDLT}, which can
be used to reduce the calculation by a factor of two.
To illustrate the evaluation of the integrals \eqref{eq:DIDLT}, we
present concrete expressions for the approximating sums when $N=4$ (which is the setting that will be implemented
later on). Defining $$\theta^k_n =k\pi/M\ \text{ and }\
r_n=\frac{2M}{5T_n},$$ we obtain {\small
\begin{eqnarray*}
&&\hspace{-0.8cm}DID(S_0,h,\textbf{T})\approx\frac{2}{5^4 T_1 T_2 T_3 T_4 }\sum_{k_1=0}^{M-1} \sum_{k_2=0}^{M-1} \sum_{k_3=0}^{M-1} \sum_{k_4=0}^{M-1}\\
&& \beta_{k_1} \beta_{k_2} \beta_{k_3}\beta_{k_4} f(q_{k_1}/T_1,q_{k_2}/T_2,q_{k_3}/T_3,q_{k_4}/T_4)+\overline{\beta}_{k_1} \beta_{k_2} \beta_{k_3}\beta_{k_4} f(\overline{q}_{k_1}/T_1,q_{k_2}/T_2,q_{k_3}/T_3,q_{k_4}/T_4)\\
&+&\beta_{k_1} \overline{\beta}_{k_2} \beta_{k_3}\beta_{k_4} f(q_{k_1}/T_1,\overline{q}_{k_2}/T_2,q_{k_3}/T_3,q_{k_4}/T_4)+\beta_{k_1} \beta_{k_2} \overline{\beta}_{k_3}\beta_{k_4} f(q_{k_1}/T_1,q_{k_2}/T_2,\overline{q}_{k_3}/T_3,q_{k_4}/T_4)\\
&+&\beta_{k_1} \beta_{k_2} \beta_{k_3}\overline{\beta}_{k_4} f(q_{k_1}/T_1,q_{k_2}/T_2,q_{k_3}/T_3,\overline{q}_{k_4}/T_4)+\overline{\beta}_{k_1} \overline{\beta}_{k_2} \beta_{k_3}\beta_{k_4} f(\overline{q}_{k_1}/T_1,\overline{q}_{k_2}/T_2,q_{k_3}/T_3,q_{k_4}/T_4)\\
&+&\beta_{k_1} \overline{\beta}_{k_2}
\beta_{k_3}\overline{\beta}_{k_4}
f(q_{k_1}/T_1,\overline{q}_{k_2}/T_2,q_{k_3}/T_3,\overline{q}_{k_4}/T_4)+\overline{\beta}_{k_1}
\overline{\beta}_{k_2} \overline{\beta}_{k_3}\beta_{k_4}
f(\overline{q}_{k_1}/T_1,\overline{q}_{k_2}/T_2,\overline{q}_{k_3}/T_3,q_{k_4}/T_4),
\end{eqnarray*}}
where $f$ is equal to $\WH{DID}$. The weights and the nodes are given by
\begin{eqnarray*}
    q_0&=&\frac{2M}{5},\quad q_k=\frac{2k\pi}{5}(\cot(k\pi/M)+\mathbf i), \quad \quad 0<k<M,\\
    \beta_0&=&0.5 e^{q_0}, \quad \beta_k=(1+\mathbf i(k\pi/M)(1+[\cot(k \pi/M)]^2)-\mathbf i \cot(k\pi/M))e^{q_k}.
\end{eqnarray*}
Since the weights and nodes are independent of the transform, the
 calculation time of the algorithm can be reduced by pre
computing and storing weights and nodes. The speed of convergence
and the accuracy  of the Talbot algorithm will depend on the
regularity of the Laplace transform $f$. Although universal error
bounds are not known, Abate et al.~\cite{abate04} showed
numerically that the single parameter $M$ can be used to control
the error and can be seen as a measure for the precision. They
found after extensive numerical experiments that for a large class
of Laplace transforms the relative error is approximately
$10^{-0.6M}$. For high dimensional inversion, extra accuracy in
the inner sums may be needed to obtain a sufficient
degree of precision for the outer sums, which can be achieved by
increasing $M$.

\subsection{Fractional Fourier Transform}
To evaluate down-and-in call option prices (DIC), we invert
the Fourier-Laplace transform \eqref{eq:DIC} over log-strike
and time periods. For the inversion of the Laplace transform we again apply the
Talbot algorithm. In the case of two time periods, with
$$
f_v(q_1,q_2)=\WH{DIC}^*(S_0,h,v,\textbf{q}),
$$
we find that the Fourier transform $DIC^*$ can be approximated
by the following sums:
\begin{eqnarray*}
\lefteqn{DIC^*(S_0,h,v,\textbf{T})}\\
&\approx&\frac{1}{5^2 T_1 T_2 }\sum_{k_1=0}^{M-1}
\sum_{k_2=0}^{M-1} \le\{\beta_{k_1}
\beta_{k_2}f_v(q_{k_1}/T_1,q_{k_2}/T_2)+\overline{\beta}_{k_1}
\overline{\beta}_{k_2}
 f_v(\overline{q}_{k_1}/T_1,\overline{q}_{k_2}/T_2)\ri.\\
&\phantom{=}&\qquad\qquad\qquad\quad +\ \le.\overline{\beta}_{k_1}
\beta_{k_2}f_v(\overline{q}_{k_1}/T_1,q_{k_2}/T_2)+\beta_{k_1}
\overline{\beta}_{k_2}f_v(q_{k_1}/T_1,\overline{q}_{k_2}/T_2)\ri\}.
\end{eqnarray*}
Unlike the case of the inversion of $\WH{DID}$,
we cannot reduce the calculation time by two by
using complex conjugates, since the function $DIC^*$ is not real
valued. Down-and-in call prices are then obtained by inverting the
Fourier transform over strike:
$$
DIC(S_0,h,k, \textbf{T})=\frac{\te{-\alpha k}}{\pi}\int_0^{\infty} \te{-\mathbf
ivk}DIC^*(v)\td v,
$$
where $\alpha$ is the rate of exponential dampening. This integral
is approximated for a set of log-strikes between $(-x_0,x_0)$ as a
summation:
\begin{equation}\label{eq:DICsum}
DIC(S_0,h,k, \textbf{T}) \approx \frac{S \te{-\alpha
k}\te{-rT}}{\pi}\sum_{j=0}^{N-1}w_j\te{-\mathbf i\delta j(-x_0+k
\lambda)}DIC^*(\delta j) \delta, \quad \quad k=1\cdots N-1,
\end{equation}
where $(w_j)_{j=0}^{N-1}$ are the integration weights defined by
the trapezoidal rule with $w_0=w_{N-1}=0.5$ and $1$ otherwise,
$\lambda = 2x_0/N$ is the log-strike grid step-size
and $\delta$ is the $v$-grid step-size.
Carr and Madan \cite{carr98}
and Chourdakis \cite{chourdakis04} set $\delta=0.25$.

To have accurate prices for any strike, the log-strike grid spacing
$\lambda$ needs to be sufficiently small. A common approach is to apply directly the
 Fast Fourier Transform (FFT) and to compute the summation \eqref{eq:DICsum} on a fixed log-strike range $(-x_0,x_0)$ with $x_0=\pi/\delta$
using many points $N$. Bailey and Swarztrauber \cite{Bailey91}, \cite{Bailey94}
propose an alternative approach, and define the Fractional Fast
Fourier transform (FrFFT), which uses an arbitrary range.
Chourdakis \cite{chourdakis04} showed that the FrFFT can be used to
calculate option prices with less points without losing accuracy.
He reported
that the FrFFT is $45$ times faster than the FFT for
the calculation of European option prices. Since in our case the Fourier
transform $DIC^*$ is obtained numerically, we chose to
employ the FrFFT. We now briefly specify the
form of this algorithm in our setting, and refer for further details
to \cite{Bailey91}, \cite{Bailey94}, \cite{chourdakis04}.
The resulting sum is then given by
\begin{eqnarray*}
DIC(S_0,h,k,\textbf{T}) &\approx&\frac{S_0 \te{-(\alpha k+\mathbf i \pi k^2
\nu)}\te{-rT}}{\pi}\sum_{j=0}^{N-1}\widetilde{w}_j\te{-\pi \mathbf
i j^2 \nu}\te{-\pi \mathbf i(k-j)^2 \nu}DIC^*(\delta j) \delta,
\end{eqnarray*}
where $k=1\cdots N-1$, $\widetilde{w}_j=w_je^{\mathbf i x_0 \delta j}$ and $\nu=\delta x_0/N
\pi$. Extending this summation into a circular convolution over $2N$
yields
\begin{eqnarray*}
DIC(S_0,h,k,\textbf{T})&\approx& \frac{S_0 \te{-(\alpha k+\mathbf i\pi k^2
\nu)}\te{-rT}}{\pi}\sum_{j=0}^{2N-1} y_j z_{k-j}, \quad \quad
k=1\cdots N-1,
\end{eqnarray*}
where
$$
y_j=\widetilde{w}_j \te{-\pi \mathbf i j^2 \nu} DIC^*(\delta j)\delta,
\quad z_{j}= \te{-\pi \mathbf i j^2 \nu}, \quad j<N,\\
$$
and
$$
y_j=0, \quad z_{j}= \te{-\pi \mathbf i (j-2N)^2 \nu}, \quad j \geq N.
$$
This equation can be rewritten in terms of three discrete Fourier transforms:
\begin{eqnarray*}
DIC(k) &\approx&\frac{S_0\te{-(\alpha k+\mathbf i\pi k^2
\nu)}\te{-rT}}{\pi}F_k^{-1}(F_k(y)F_k(z)), \quad \quad k=1\cdots
N-1,
\end{eqnarray*}
with
$$
F(x)=\sum_{j=0}^{N-1}x_j \te{-2\pi \mathbf i j k/N},\quad \quad
\quad F^{-1}(x)=\sum_{j=0}^{N-1}x_j \te{2\pi \mathbf i j k/N}.
$$
Although the latter sum is computed by invoking two Fourier
transforms and one inverse Fourier transform, this approach has the advantage
of computing the option prices on a specific log-strike window
$(-x_0,x_0)$ with independent grids $\delta$ and $\lambda$
and requires less points.

\newpage{}

\end{document}